\documentclass[12pt,a4paper]{article}
\usepackage[utf8]{inputenc}
\usepackage{epsf}
\usepackage{latexsym,amssymb,euscript}
\usepackage[dvips]{graphicx}
\usepackage[numbers,sort&compress]{natbib}
\usepackage{amsmath}
\usepackage{slashed}
\usepackage{booktabs}
\usepackage{hyperref}
\usepackage{braket}
\usepackage{chngcntr}
\usepackage{bbold}
\usepackage{graphics}
\usepackage{graphicx}
\usepackage{caption}
\usepackage{subcaption}
\usepackage{pdfpages}
\usepackage[titletoc]{appendix}
\graphicspath{{./figures/}}
\hypersetup{
 linktocpage = true,
 urlcolor = purple,
 colorlinks = true,
 linkcolor = purple,
 anchorcolor = purple,
 citecolor = purple,
 pdfstartview = {XYZ null null 1.25} 
           }
\usepackage[left=2cm, right=2cm]{geometry}
\usepackage{pstricks}
\usepackage{color}
\usepackage{float}
\usepackage{academicons}
\definecolor{orcidlogocol}{HTML}{A6CE39}

\newcommand{\orcidFGC}{\href{https://orcid.org/0000-0003-3299-2203}{\includegraphics[scale=0.1]{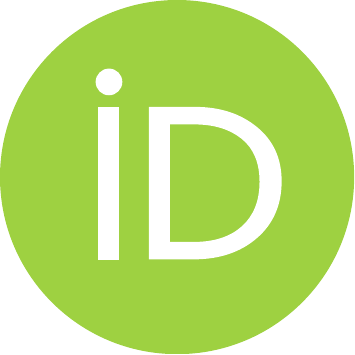}}}

\begin{document}

\begin{titlepage}

\begin{center}
  {\LARGE \bf 3D tomography of the nucleon: transverse-momentum-dependent gluon distributions}
\end{center}

\vskip 0.5cm

\centerline{
Francesco~Giovanni~Celiberto$^{\;1,2,3\;\dagger}$ \orcidFGC
}
\vskip .6cm

\centerline{${}^1$ {\sl European Centre for Theoretical Studies in Nuclear Physics and Related Areas (ECT*),}}
\centerline{\sl I-38123 Villazzano, Trento, Italy}
\vskip .2cm
\centerline{${}^2$ {\sl Fondazione Bruno Kessler (FBK), 
I-38123 Povo, Trento, Italy} }
\vskip .2cm
\centerline{${}^3$ {\sl INFN-TIFPA Trento Institute of Fundamental Physics and Applications,}}
\centerline{\sl I-38123 Povo, Trento, Italy}
\vskip 2cm

\begin{abstract}
\vspace{0.50cm}
\hrule \vspace{0.75cm}
We perform explorative analyses of the 3D gluon content of the proton via a study of (un)polarized twist-2 gluon TMDs, calculated in a spectator model for the parent nucleon. Our approach encodes a flexible parametrization for the spectator-mass density, suited to describe both moderate and small-$x$ effects. All these prospective developments are relevant in the investigation of the gluon dynamics inside nucleons and nuclei, which constitutes one of the major goals of new-generation colliding machines, as the EIC, the HL-LHC and NICA. 
\vspace{0.75cm} \hrule
\vspace{0.75cm}
{
 \setlength{\parindent}{0pt}
}
\end{abstract}

\vfill
$^{\dagger}${\it e-mail}:
\href{mailto:fceliberto@ectstar.eu}{fceliberto@ectstar.eu}

\end{titlepage}

\section{Introductory remarks}
\label{introd}

The search for clues of New Physics is a primary target of current and upcoming studies at the Large Hadron Collider (LHC) and at new-generation hadron, lepton and lepton-hadron colliders. This is the best moment to deepen our understanding of strong interactions and, notably, of the hadron structure in terms of parton distributions.
Although significant results were obtained concerning quark transverse-momentum dependent distribution functions (TMDs), the deep knowledge on their formal properties being surrounded by a rich and wealthy phenomenology, the gluon-TMD field represents an almost uncharted territory.

Gluon TMD densities were listed for the first time in ref.~\cite{Mulders:2000sh} and then classified in refs.~\cite{Meissner:2007rx,Lorce:2013pza,Boer:2016xqr}.
First phenomenological analyses of the unpolarized ($f_1^g$) and  Sivers ($f_{1T}^{\perp g}$) gluon-TMD functions were done in refs.~\cite{Lansberg:2017dzg,Gutierrez-Reyes:2019rug,Scarpa:2019fol} and~\cite{Adolph:2017pgv,DAlesio:2017rzj,DAlesio:2018rnv,DAlesio:2019qpk}, respectively.
From the formal point of view, it is widely recognized that different classes of processes probe distinct 
gauge-link structures, thus resulting in a more diversified kind of modified universality with respect to the quark case.
This brings us to a distinction between $f\text{-type}$ and $d\text{-type}$ gluon TMDs, also known in the context of small-$x$ studies as Weisz\"acker--Williams and dipole TMDs~\cite{Kharzeev:2003wz,Dominguez:2010xd,Dominguez:2011wm}.
The $[+,+]$ and $[-,-]$ gauge links emerge in $f$-type TMDs, whereas the $[+,-]$ and $[-,+]$ gauge links are the building blocks of $d$-type ones.
More intricate, box-loop gauge links appear in processes where multiple color states are present both in the initial and final state~\cite{Bomhof:2006dp}, thus leading, however, to factorization-breaking effects~\cite{Rogers:2013zha}.
At large transverse momentum and at small-$x$, the unpolarized and linearly polarized gluon TMDs, $f_1^g$ and $h_1^{\perp g}$, are linked~\cite{Dominguez:2011wm} to the unintegrated gluon distribution (UGD), originally defined in the BFKL formalism~\cite{Fadin:1975cb,Kuraev:1976ge,Kuraev:1977fs,Balitsky:1978ic} (see Refs.~\cite{Hentschinski:2012kr,Besse:2013muy,Bolognino:2018rhb,Bolognino:2018mlw,Bolognino:2019bko,Bolognino:2019pba,Celiberto:2019slj,Brzeminski:2016lwh,Garcia:2019tne,Celiberto:2018muu,Celiberto:2020wpk,Bolognino:2019yls,Celiberto:2020tmb,Celiberto:2020rxb} for recent applications).


With the aim of fulfilling the need for a flexible model suited to phenomenology, we present a common framework for all $T$-even and gluon TMDs at twist-2, calculated in a spectator model for the parent nucleon and encoding effective small-$x$ effects from the BFKL resummation. At variance with respect to previous works, our approach embodies a flexible parametrization for the spectator-mass spectral density, allowing us to improve the description in a wide range of $x$.

\section{A spectator-model way to gluon TMD distributions}
\label{theory}

We have extended spectator-model calculations, done for quark TMDs~\cite{Bacchetta:2008af,Bacchetta:2010si}, to the case of unpolarized and polarized $T$-even gluon TMDs at twist-2. 
The assumption of the model is that a nucleon (a proton, in our study) can emit a gluon with longitudinal-momentum fraction $x$ and transverse momentum $\boldsymbol{p}_T$, while the remainders are treated as a single spectator particle. The nucleon-gluon-spectator coupling is described by an effective vertex containing two form factors, given as dipolar functions of $\boldsymbol{p}_T^2$. Dipolar expressions are very useful, since they permit us to cancel the gluon-propagator singularity, dampen the effects of high transverse momenta where the TMD formalism cannot be employed, and quench logarithmic divergences emerging in $\boldsymbol{p}_T$-integrated densities.

We have enhanced the standard spectator-model description by allowing the spectator mass, $M_X$, to be in a range of values weighed by a spectral function. We fit model parameters by reproducing the gluon unpolarized and helicity collinear parton distribution functions (PDFs), obtained in global fits, at the initial scale, $Q_0 = 1.64$ GeV. A detailed description of our model as well as technical aspects of our fit are given in ref.~\cite{Bacchetta:2020vty}.

\begin{figure}[tb]
 \centering
 \includegraphics[scale=0.25,clip]{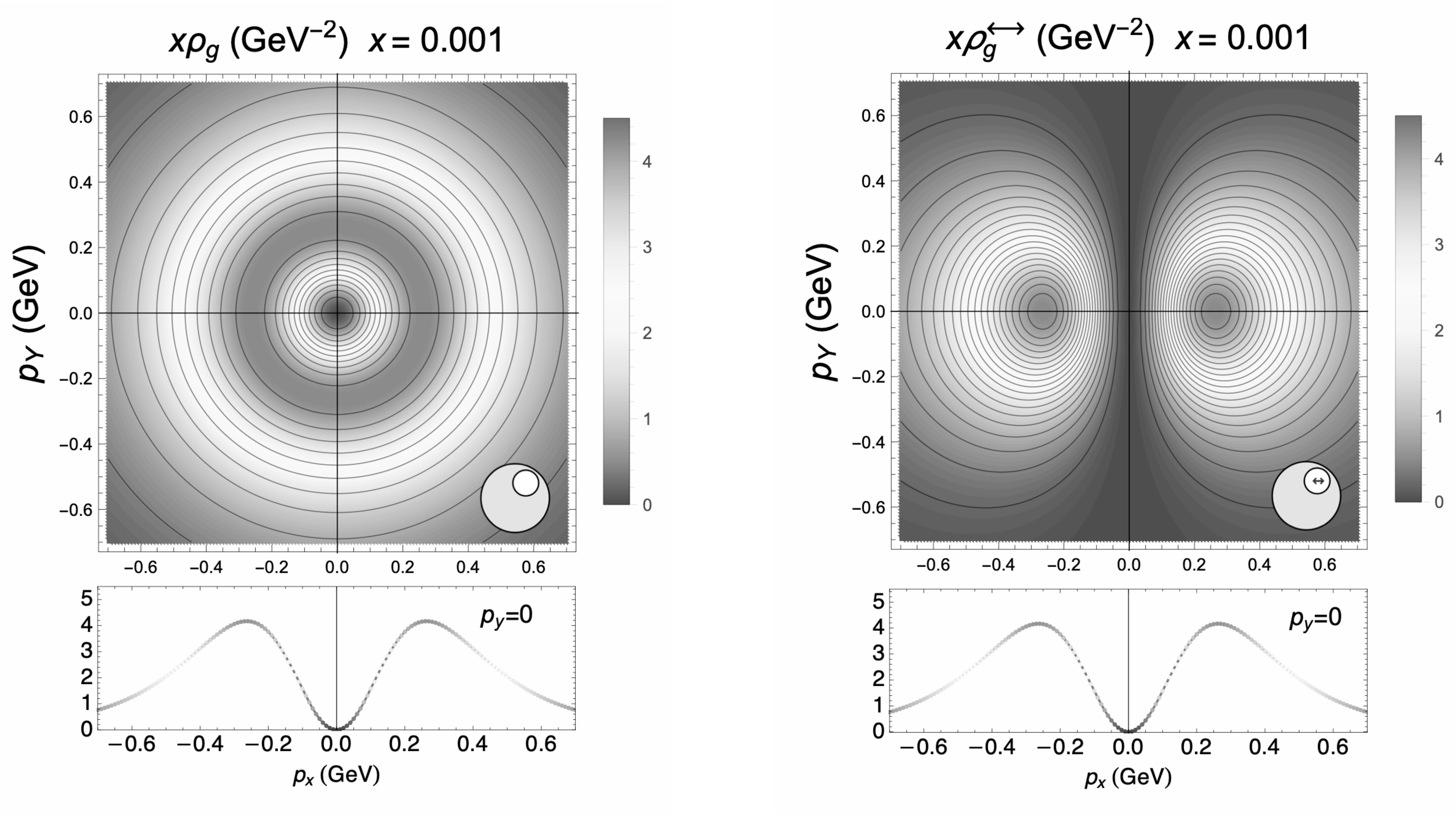}

 \caption{Tomographic imaging of the unpolarized (left) and Boer--Mulders (right) gluon TMD densities as functions of $\boldsymbol{p}_T \equiv (p_x, p_y)$, for $x = 10^{-3}$ and at the initial energy scale, $Q_0 = 1.64$ GeV. Ancillary 1D plots show the density at $p_y = 0$.}
 \label{fig:f1_h1p}
\end{figure}

%
We present a selection of results of our gluon TMDs for different combinations of nucleon spin and emitted-gluon polarization. In particular, we focus on the tomographic imaging in the $\boldsymbol{p}_T$-plane of the so-called $\rho$-densities. For an unpolarized nucleon with mass $M$, we define the unpolarized density
\begin{equation}
 \label{rho_unpol}
 x \rho (x, p_x, p_y) \equiv x f_1^g (x, \boldsymbol{p}_T^2)
\end{equation}
as the probability distribution of finding an unpolarized gluon at given $x$ and $\boldsymbol{p}_T$. Analogously, we identify the Boer--Mulders density
\begin{equation}
 \label{rho_BM}
  x \rho^{\leftrightarrow} (x, p_x, p_y) \equiv \frac{1}{2} \left[ x f_1^g (x, \boldsymbol{p}_T^2) - \frac{p_y^2 - p_x^2}{2 M^2} \, x h_1^{\perp g} (x, \boldsymbol{p}_T^2) \right]
\end{equation}
as the probability of finding a gluon linearly polarized in the transverse plane, as a function of $x$ and $\boldsymbol{p}_T$.
Contour plots in Fig.~\ref{fig:f1_h1p} show the $\boldsymbol{p}_T$-behavior of $x \rho (x, p_x, p_y)$ (left panel) and x $\rho^{\leftrightarrow} (x, p_x, p_y)$ (right panel) for $x = 10^{-3}$, the color code identifying the size of the oscillation of each density along the $p_{x,y}$ directions. Ancillary 1D panels below each contour plot ease the visualization of these oscillations on the $p_x$ axis, at $p_y = 0$. Since $x \rho (x, p_x, p_y)$ describes unpolarized gluons inside an unpolarized proton, it exhibits a cylindrical symmetry around the direction of motion of the nucleon pointing towards the reader. Conversely, the description of linearly polarized gluons inside an unpolarized nucleon leads to a dipolar shape for $x \rho^{\leftrightarrow} (x, p_x, p_y)$. 

\section{Closing statements}
\label{conclusions}

We have developed a consistent framework, based on a spectator-model description, where all the leading-twist $T$-even gluon distributions are concurrently generated. Moderate and small-$x$ effects are embodied in our densities by weighing the spectator-system mass via a flexible spectral function. At the present stage, our model is valid for both the $f$-type and $d$-type gluon TMDs. An extension to the case of leading-twist $T$-odd gluon TMDs is underway.
Our results for the tomography in the momentum space of (un)polarized gluons inside (un)polarized protons can represent a useful guidance on the exploration of observables sensitive to gluon-TMD dynamics at new-generation colliders, as the \emph{Electron-Ion Collider}~(EIC)~\cite{EICUGYR:2020}, the \emph{High-Luminosity Large Hadron Collider} (HL-LHC)~\cite{Chapon:2020heu}, and NICA~\cite{Arbuzov:2020cqg}.

\section*{Acknowledgements}

The author acknowledges support from the Italian Ministry of Education, Universities and Research under the FARE grant ``3DGLUE'' (n. R16XKPHL3N), and from the INFN/NINPHA project.
The author thanks A. Bacchetta, M. Radici, and P. Taels for collaboration and stimulating discussions on the subject of this work, and the Universit\`a degli Studi di Pavia for the warm hospitality.

\bibliographystyle{apsrev}
\bibliography{references}

\end{document}